\documentclass{aa}
\usepackage{psfig,natbib}

\def\chandra{{\it Chandra}} 
\def\xmm{{\it XMM-Newton}} 
\def\asca{{\it ASCA}} 
\def\hst{{\it HST}} 
 
\def\sax{{\it BeppoSAX}} 
 
\def\rosat{{\it ROSAT}}

\def\nh{cm$^{-2}$}
\def\arcsec{$^{\prime\prime}$}
\def\arcmin{$^{\prime}$}
\def\deg{$^{\circ}$}
\def\ltsima{$\; \buildrel < \over \sim \;$}
\def\simlt{\lower.5ex\hbox{\ltsima}} 
\def\gtsima{$\; \buildrel > \over \sim \;$}
\def\simgt{\lower.5ex\hbox{\gtsima}} 

\begin{document}

\title{On the origin of the X-rays and the nature of accretion in \object{NGC~4261}}
\author{M. Gliozzi\inst{1}  
\and  R.M. Sambruna\inst{1} 
\and W.N. Brandt\inst{2}} 
\offprints{mario@physics.gmu.edu} 
\institute{George Mason University, Dept. of Physics \&
Astronomy \& School of Computational Sciences, 4400 University Drive, MS 3F3, Fairfax, VA 22030
\and  The Pennsylvania State University, Department of
Astronomy \& Astrophysics, 525 Davey Lab, University Park, PA 16802}

\date{Received: ; accepted: }

\abstract{We report on the X-ray properties of the radio galaxy
\object{NGC~4261}, combining information from the \xmm, \chandra, and
\sax\ satellites. Goals of this study are to investigate the origin 
of the X-rays from this low-power radio galaxy and the nature of the 
accretion process onto the central black hole. The X-ray spectrum of the 
nuclear source extending up to 100--150 keV is well described by a
partially covered (covering factor $>$ 0.8) power law with a photon
index $\Gamma\simeq 1.5$ absorbed by a column density $N_{\rm H}>
5\times 10^{22}{~\rm cm^{-2}}$. The X-ray luminosity
associated with the non-thermal component is $\sim 5 \times 10^{41} {~\rm erg~s^{-1}}$.
The nuclear source is embedded in a diffuse hot gas ($kT\sim0.6-0.65$ keV), whose
density profile implies a Bondi accretion rate of $\sim 4.5\times 10^{-2} 
{~\rm M_\odot~yr^{-1}}$. For the first time rapid X-ray variability is detected in a 
low-power radio galaxy at more than 99\% confidence level.
The observed X-ray spectral and variability properties
indicate the accretion flow as the most likely origin of the bulk X-ray continuum.
This conclusion is strengthened by energetic considerations based 
on a comparison between the X-ray luminosity and the kinetic power of the jet, which
also suggest that the Bondi accretion rate overestimates the actual accretion rate onto 
the black hole.
}
\titlerunning{Origin of X-rays and nature of accretion in \object{NGC~4261}}
\authorrunning{M.~Gliozzi et al.}
\maketitle

\section{Introduction}
One of the most interesting issues of extragalactic astrophysics is the
condition of matter near black holes. The importance of this issue
is stressed by the recent discovery from HST and ground-based
observations that most nearby galaxies harbor supermassive black holes
(e.g., Gebhardt et al. 2000, Ferrarese et al. 2001).
Together with independent
evidence that weak nuclear activity is common in both spirals and
ellipticals (Low-Ionization Nuclear Emission-Line Regions, 
hereafter LINERs; e.g., Heckman 1980), 
this underlines a fundamental link between ``normal" and
active galaxies. Since accreting supermassive
black holes are widely believed to be responsible for the nuclear
activity in active galactic nuclei (AGN), one naturally wonders why
intense activity is limited to a minority of galaxies.  
Inefficient accretion onto the central black hole is generally considered as
the most likely explanation. 
This could be due either to
radiatively inefficient advection (or convection) dominated accretion
flows (e.g., Narayan 2002), or to unsteady feedback-modulated
accretion (e.g., Binney \& Tabor 1995, Ciotti \& Ostriker 2001), 
which can stop or decrease
to a low duty cycle the accretion onto the massive black hole. 

Weak AGN may represent the link between powerful AGNs and ``normal"
galaxies. Therefore, an important role in understanding the nature of
accretion around massive black holes can be played by the the study of
weak AGNs at X-ray energies, where the combination of spatial,
spectral, and temporal analyses offer an ideal means by which we can
understand the physics of their central engines and in particular the
accretion process onto their black holes.

The nearby ($z$=0.0074) giant elliptical \object{NGC~4261} (3C270)
contains a supermassive black hole with $M_{\rm BH}=
(4.9\pm1.0)\times10^8 M_{\odot}$ (Ferrarese, Ford, \& Jaffe 1996). It is one of
a handful of nearby early-type galaxies where low-luminosity nuclear activity
is detected. It is classified as a Fanaroff-Riley I (FR~I) 
radio galaxy, with a compact core and
twin jets extending E-W, oriented at $\theta=(63 \pm 3)$\deg\ with respect
to the line of sight (Piner, Jones, \& Wehrle 2001). Optical
spectroscopy reveals a 
LINER in the nucleus of the galaxy (Ho, Filippenko, \& Sargent
1995).  
\hst/STIS ultraviolet images of \object{NGC~4261} show the
presence of a jet-like structure on pc scales with position angle
consistent with the orientation of the radio and X-ray jets (Chiaberge et al. 2003). 
In the X-ray band, diffuse thermal emission 
plus a non-thermal nuclear component
is present at low energies in \rosat\ images (Worrall \&
Birkinshaw 1994). More recently, \chandra\ spectral results have been
reported by Chiaberge et al. (2003), showing a continuum described by
a flat power law with photon index $\Gamma \sim 1.4$
and an excess
column density of N$_H \sim 6 \times 10^{22}$ \nh. Marginal evidence for
an unresolved Fe line was previously claimed from a 40 ks
\asca\ exposure 
(Sambruna, Eracleous, \&
Mushotzky 1999). These spectral properties and the presence of an unresolved 
Fe K line at 6.9 keV have been confirmed by
Sambruna et al. (2003; hereafter Paper I) using \xmm\ data, which also
show evidence for significant variability on short time scales.

\begin{table*} 
\caption{X-ray observation log}
\begin{center}
\begin{tabular}{lllllc}
\hline
\hline
\noalign{\smallskip}
Satellite & Instrument & Observation Date & Exposure & Count Rate & Extraction Radius\\
~ & &[yy/mm/dd] & [ks]& $[\rm s^{-1}]$ \\
\noalign{\smallskip}       
\hline
\noalign{\smallskip}
\noalign{\smallskip}
\chandra\ & ACIS-S3 & 00/05/06 & 32.0$^{\rm a}$ & $(7.04\pm0.15)\times10^{-2}$ & 2\arcsec\\
\noalign{\smallskip}
\hline
\noalign{\smallskip}
\sax\  & MECS23 & 00/12/27 & 67.4 & $(8.51\pm0.39)\times10^{-3}$ & 2\arcmin\\
\noalign{\smallskip}
\hline
\noalign{\smallskip}
\xmm\  & EPIC pn & 01/12/16 & 21.3$^{\rm a}$ & $(3.03\pm0.04)\times10^{-1}$ & 20\arcsec\\
\noalign{\smallskip}
\hline
\end{tabular}
\smallskip
\smallskip
\end{center}
$^{\rm a}$ Effective exposure  after excluding times of background flares.
\label{table:log}
\end{table*}

In this paper, we present a detailed study of the X-ray nuclear and
circumnuclear properties of \object{NGC~4261}, combining the complementary
capabilities of \xmm, \chandra, and \sax.  
The main questions addressed in this study are 
1) What is the origin of the X-ray
emission from this galaxy? In particular, concerning the nuclear
emission, are the bulk of the X-rays due to the inner jet or is the
accretion process responsible for them? 2) What is the nature of
the accretion process at work in the nucleus of \object{NGC~4261}?  

In $\S2$ we describe the 
observations and data reduction. In $\S3$ we perform a spatial analysis
of the circumnuclear region utilizing \chandra\ data. In $\S4$ and $\S5$
we carry out temporal and spectral analyses, respectively. In $\S6$
we discuss the nuclear properties inferred from the X-ray data and compare
them with the jet properties on pc scales derived from radio observations. Finally, in 
$\S7$, we summarize the main conclusions. Throughout the paper we use a Friedman 
cosmology with $H_0=75~{\rm km~s^{-1}~Mpc^{-1}~and}~q_0=0.5$. With
these cosmological parameters, 1\arcsec=141 pc at the distance of \object{NGC~4261}. 

\section{Observations and data reduction}

Table~\ref{table:log}  summarizes the X-ray observations of \object{NGC~4261} carried
out with \chandra, \sax, and \xmm. The
background-subtracted count rates are given in column 5. 
Details are given below for each satellite. 

\subsection{XMM-Newton} 

We observed \object{NGC~4261} with \xmm\ on 2001 October 16
for a duration of $\sim 27$ ks with the EPIC pn, and
for $\sim 32$ ks with EPIC MOS1 and MOS2. All of the EPIC cameras were
operated in full-frame mode with a medium filter. The recorded events
were reprocessed and screened with the latest available release of the
\xmm\ Science Analysis Software (\verb+SAS+ 5.3.3) to remove known hot
pixels and other data flagged as bad; only data with {\tt FLAG=0} were
used. Investigation of the full--field light curves revealed the
presence of two periods of background flaring; these events were
screened reducing the effective total exposures to $\sim 21$ ks for the
EPIC pn and $\sim 26$ ks for the MOS cameras.
Background data were extracted from source-free circular regions
on the same chips containing the source.  There
are no signs of pile-up in the pn or MOS cameras
according to the {\tt SAS} task {\tt epatplot}.  With an
extraction radius of 20\arcsec\ the detected count rates in the energy
range 0.3--10 keV are $(9.52\pm0.20)\times10^{-2}{\rm~s^{-1}}$ for the
MOS1, $(8.91\pm0.20)\times10^{-2}{\rm~s^{-1}}$ for the MOS2, and
$(3.03\pm0.04)\times10^{-1}{\rm~s^{-1}}$ for the pn.

The spectra were rebinned such that
each spectral bin contains at least 20 counts in order to apply
$\chi^2$ minimization, and fitted using the {\tt XSPEC v.11} software
package (Arnaud 1996). The quoted errors on the derived best-fitting
model parameters correspond to a 90\% confidence level for one
interesting parameter (i.e., a $\Delta\chi^2=2.7$ criterion) unless
otherwise stated.  The latest publicly available responses were used.
The same procedure was applied to the \chandra\ and \sax\ data.

\subsection{Chandra} 

\object{NGC~4261} was observed with \chandra\ on 2000 May 6.
The observations
were carried out in a subarray mode (frametime $\sim$ 1.8 s) to
minimize the possible pile-up of the nucleus. The data reduction was
performed with \verb+CIAO+ v. 2.3 and \verb+CALDB+ v. 2.18. The data
were reprocessed with {\tt acis\_process\_events} and screened to
exclude periods of flaring background, reducing the effective exposure
to $\sim 32$ ks.
In order to take
into account the continuous degradation of the ACIS quantum efficiency
(QE) due to molecular contamination of the optical blocking filters,
we applied {\tt ACISABS} \footnote[1]{see the Chartas \& Getman model description at
{\tt www.astro.psu.edu/users/chartas/xcontdir/xcont.html}} to
the ancillary response function (ARF) file.
The
S3 background-subtracted count rate in the 0.35--9 keV energy band is 
$(7.04\pm0.15)\times10^{-2} {~\rm cts~s^{-1}}$. 
  
\subsection{BeppoSAX} 

\sax\ observed \object{NGC~4261} with the
Narrow Field Instruments, LECS (0.1--4.5 keV) and MECS
(1.3--10 keV), and with the PDS (15--220 keV) on 2000 December 27, with 
effective exposures of 19.6 ks, 67.4 ks, and  32.6 ks, respectively. 
The LECS and MECS are imaging instruments, while the PDS is a phoswich detector system.
The observations were performed with 2 active MECS units. 
Standard data reduction techniques were employed, following the prescription
given by Fiore et al. (1999). LECS and MECS spectra and light curves were
extracted from regions with radii of 4\arcmin\ and 2\arcmin, respectively, in
order to maximize the accumulated  counts at both low and high energies.
An inspection of the MECS f.o.v. argues against the presence of confusing
bright sources in the PDS data.
Background spectra were extracted from high Galactic latitude ``blank" fields,
whereas for the light curves a source-free region in the field of view was
used. The background-subtracted count rates are $(9.9\pm0.8)\times10^{-3}
{~\rm cts~s^{-1}}$ for the LECS in the 0.1--4.5 keV band,
$(8.5\pm0.4)\times10^{-3}{~\rm cts~s^{-1}}$  for the MECS in the 2--10 keV band, 
and $(6.9\pm3.6)\times10^{-2}{~\rm cts~s^{-1}}$ for the PDS in the 15--150 keV
energy band. 

\section{Spatial Analysis}

\begin{figure}
\noindent{\psfig{figure=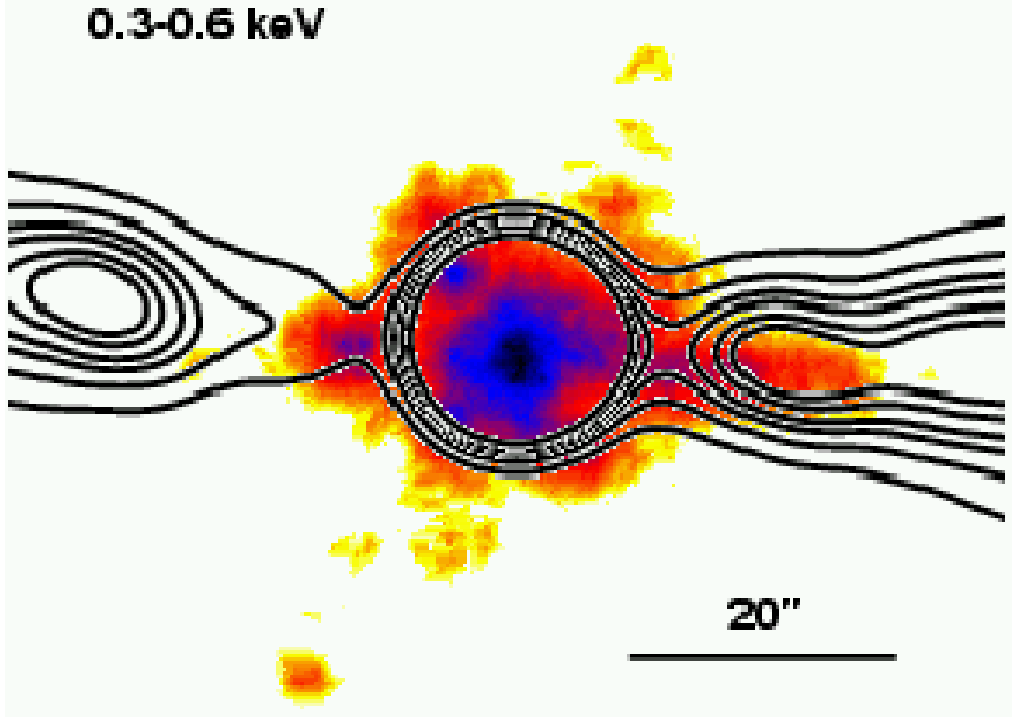,height=3.4cm,width=4.3cm,%
bbllx=160pt,bblly=282pt,bburx=452pt,bbury=509pt,angle=0,clip=}}
{\psfig{figure=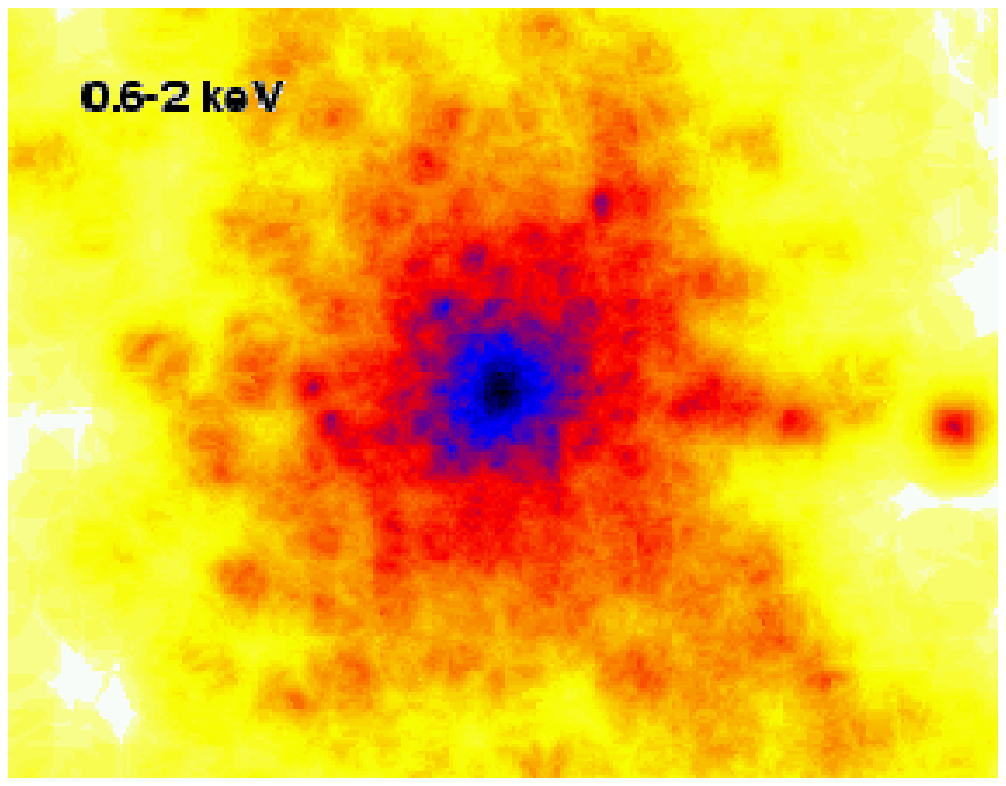,height=3.4cm,width=4.3cm,%
bbllx=160pt,bblly=282pt,bburx=452pt,bbury=509pt,angle=0,clip=}}
\noindent{\psfig{figure=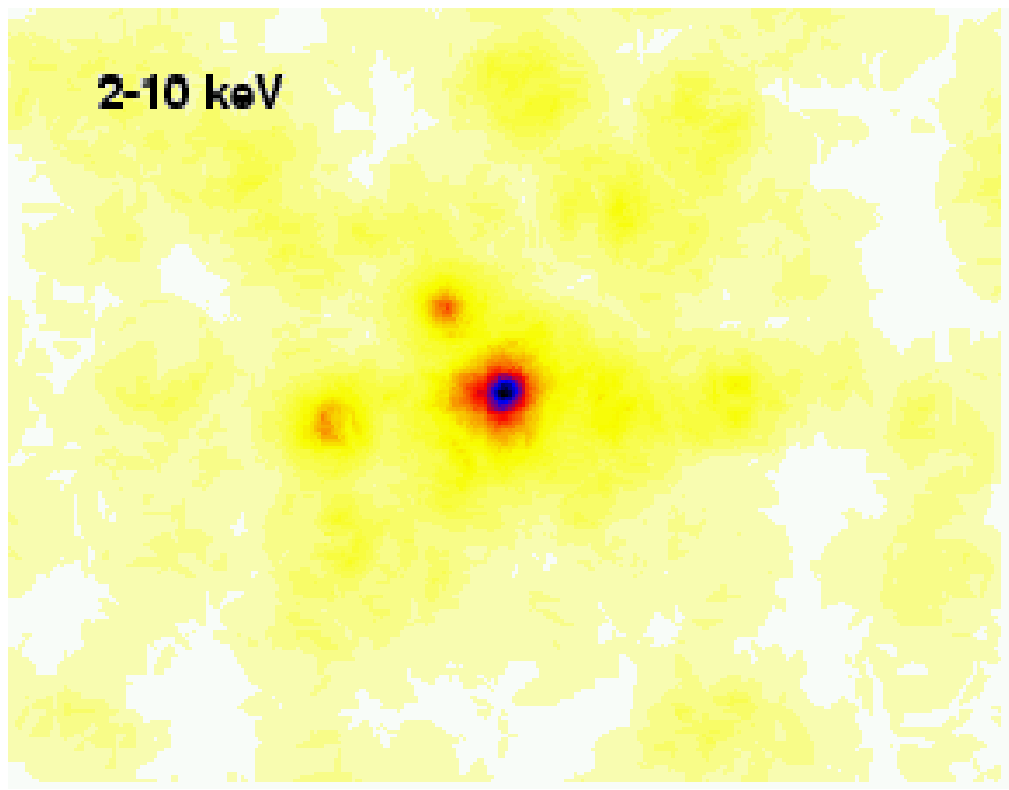,height=3.4cm,width=4.3cm,%
bbllx=160pt,bblly=282pt,bburx=452pt,bbury=509pt,angle=0,clip=}}
{\psfig{figure=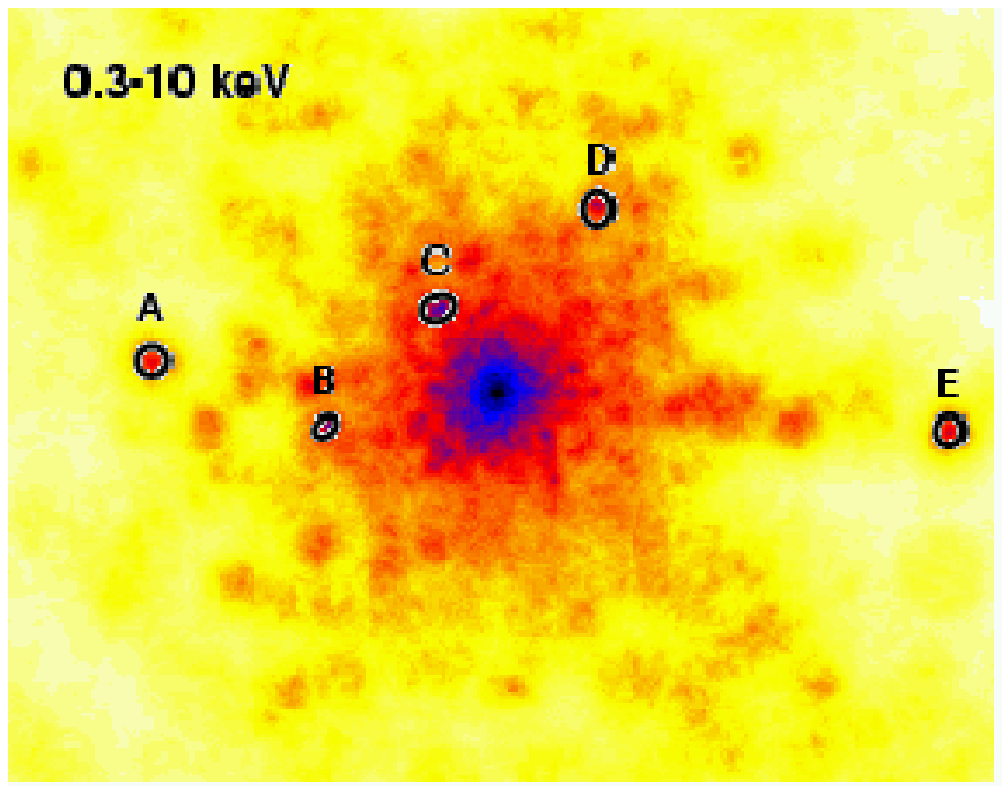,height=3.4cm,width=4.3cm,%
bbllx=160pt,bblly=282pt,bburx=452pt,bbury=509pt,angle=0,clip=}}
\caption{\chandra\ ACIS-S3 adaptively smoothed images of
\object{NGC~4261} in different energy bands. The top left panel shows
the emission in the ultrasoft range 0.3--0.6 keV, with VLA radio
contours overlaid.
The top right and bottom left panels display the soft (0.6--2 keV)
and hard (2--10 keV) X-ray emission, respectively.
In the bottom right panel
the total emission (0.3--10 keV) is shown; the
sources detected with {\tt wavdetect} are marked.
At the distance of \object{NGC~4261}, 20\arcsec\ corresponds to a
distance of 2.8 kpc. North is up, and East to the left.}
\label{figure:cha-ima}
\end{figure}

With its subarcsecond spatial resolution, \chandra\ provides us with
the opportunity to take a direct look at the faint nuclear emission
and to disentangle the different X-ray components in
NGC~4261. Figure~\ref{figure:cha-ima} shows four adaptively smoothed
images of \object{NGC~4261} in different energy bands as seen by
ACIS-S3. In the ultrasoft range (0.3--0.6 keV, top left panel), the
X-ray emission is preferentially distributed along the jet direction
indicated by the VLA radio contours superimposed. It is worth noticing
that \chandra\ detects an X-ray jet on kpc scales, i.e., on scales larger
than the jet-like feature observed in the UV ($\sim 20\times 60$ pc;
Chiaberge et al. 2003), and much larger than the VLBA radio jet
($\sim 1\times 3$ pc; Piner et al. 2001). 

In the 0.6--2 keV
range (top right panel), the radiation is uniformly distributed on kpc
scales (20\arcsec\ corresponds to a distance of 2.8 kpc). The hard
X-rays (2--10 keV, bottom left panel) are consistent with a point-like
source. Finally, several point-like sources (bottom right panel) are
found with {\tt wavdetect} in the circumnuclear region of
\object{NGC~4261}. 
Details about source detection and the serendipitous sources' properties are
reported in the Appendix.
Three of these sources are located within 20\arcsec\ of the nucleus of \object{NGC~4261}
and therefore fall in the \xmm\ extraction region. However, their contribution to the 
X-ray flux is negligible: the brightest source
(source C) contributes only $\sim3\%$ of the total counts,
while the remaining sources contribute less than $1\%$. 

A more quantitative method to derive information from the photon
spatial distribution is to extract a radial
profile.
Using the {\tt ciao} tools {\tt dmextract} and {\tt dmtcalc}, we
extracted the azimuthally-averaged radial profile of the
circumnuclear region out to 100\arcsec\ (corresponding to $\sim$ 14
kpc). A series of annular regions, with an increment of the radius of
2\arcsec, was used to extract the non-background-subtracted radial
profile in the 0.3--2 keV range.

\begin{figure}
{\psfig{figure=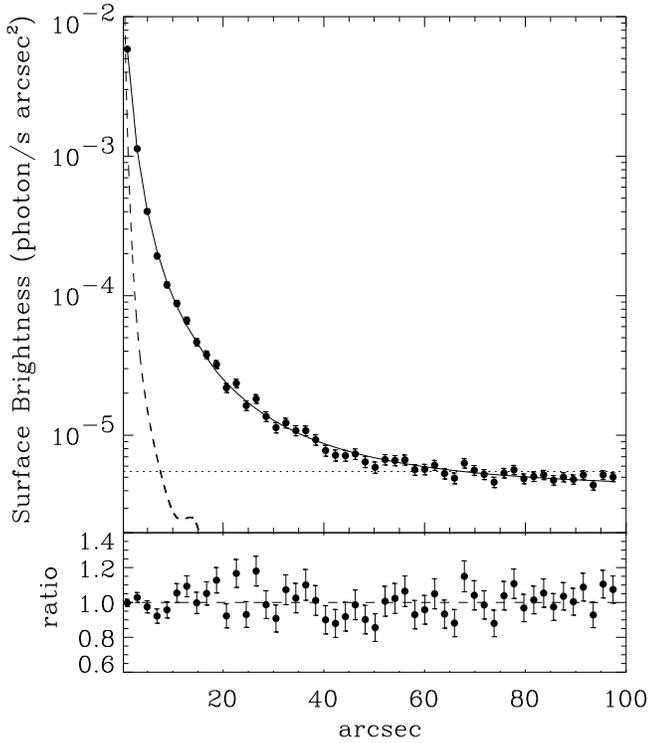,height=10cm,width=8.7cm,%
bbllx=55pt,bblly=60pt,bburx=455pt,bbury=520pt,angle=0,clip=}
\caption{\chandra\ ACIS-S3 azimuthally averaged surface brightness profile
of \object{NGC~4261}. The solid line represents the $\beta$-model
which best-fits the data, while the dashed and dotted lines represent
the PSF and the background, respectively. The lower panel shows the
data-to-model ratio. At the distance of \object{NGC~4261}, 100\arcsec\ 
corresponds to a distance of $\sim$
14 kpc.} 
\label{figure:radprof2}}
\end{figure}

To determine the physical properties
of the extended X-ray emission, we fitted the surface--brightness
profile with a $\beta$-model (e.g., Cavaliere \& Fusco-Femiano 1976) of
the form

\begin{equation}
S(r)=S_0\left(1+{r^2\over r_c^2}\right)^{-3\beta+1/2}
\end{equation}
The energy-dependent PSF (see Donato et al. 2003 for details on its derivation)
and the background were explicitly included in the
fit. The result is shown in Figure \ref{figure:radprof2}.

The best-fit values ($\chi^2_{\rm red}=1.2$, 46 d.o.f.) for the
$\beta$-model are $S_0=(6.5\pm0.3)\times10^{-3}{~\rm
cts~s^{-1}~arcsec^{-2}}$, $\beta=0.53\pm0.01$, and
$r_c=(1.40\pm0.06)$\arcsec. The quoted errors are 1$\sigma$.
The value of $\beta$ is consistent
with typical values for gas confined in nearby elliptical galaxies
(e.g., Forman et al. 1985). On the other hand, the core radius is considerably
smaller than the value inferred from 
\rosat\ PSPC data by Worrall \& Birkinshaw (1994) with $\beta$ fixed
at 2/3.  This is likely due to the different spatial resolutions
of the two instruments: the \chandra\ ACIS-S with its subarcsecond
resolution can resolve emission on very small scales, whereas the
\rosat\ PSPC with a PSF FWHM of $\sim$ 25\arcsec\ can only detect
extended emission on group scales.

The physical parameters of the extended emission can be obtained by de-projecting
the surface-brightness profile. With this method we can derive the corresponding
density profile (see, e.g., Ettori 2000):

\begin{equation}
n(r)=n_0\left(1+{r^2\over r_c^2}\right)^{-{3\beta\over 2}}
\end{equation}
This relation assumes isothermal, hydrostatic equilibrium in spherical symmetry.
We adopted the cooling function value (e.g., Sarazin 1988) for
the hot gas temperature and the abundances derived from the spectral
analysis.  The resulting central particle density is  
$n_0=0.17\pm0.01~\rm{cm}^{-3}$. 

\section{Temporal Analysis}
\begin{table}[ht] 
\caption{Short-term X-ray variability of \object{NGC~4261}}
\begin{center}
\begin{tabular}{lllll}
\hline
\hline
\noalign{\smallskip}
Energy band &  Radius   & $\chi^2_{\rm red}$ (22 d.o.f.)&  $P_{\chi^2}$ & 
$F_{\rm var}$ ${\rm ^a}$ (\%)
\\ 
\noalign{\smallskip}       
\hline
\noalign{\smallskip}
\noalign{\smallskip}
  & 10\arcsec & 1.74 & 1.2\% & $6.7\pm2.2$   \\
\noalign{\smallskip}
0.3--10 keV  & 20\arcsec & 1.85 & 0.6\% & $5.7\pm1.8$   \\
\noalign{\smallskip}
  & 30\arcsec & 1.60 & 2\% & $4.4\pm1.7$   \\
\noalign{\smallskip}
\hline
\noalign{\smallskip}
\noalign{\smallskip}
0.3--0.8 keV  & 20\arcsec & 2.03 & 0.2\% & $8.9\pm2.7$   \\
\noalign{\smallskip}
0.8--2 keV    & 20\arcsec & 1.03 & 36\% & $3.0\pm4.4$   \\
\noalign{\smallskip}
2--10 keV     & 20\arcsec & 1.14 & 24\% & $6.8\pm6.0$   \\
\noalign{\smallskip}
\hline
\end{tabular}
\end{center}
${\rm ^a}$ The errors on $F_{\rm var}$ are calculated as in
Edelson et al. (2002) and should be considered conservative estimates
of the true uncertainty. 
\label{table:shortvar}
\end{table}
As shown in Paper I, rapid variability is present in the nucleus
of  \object{NGC~4261}.
To investigate further the short-term variability  
only EPIC pn data are useful. The reason why MOS data cannot increase
the photon statistics is the following: the MOS cameras
share with the Reflection Grating Spectrometers (RGS1 and RGS2,
respectively) the focal planes of their respective X-ray telescopes.
As a consequence, they suffer from low photon statistics and their
intensity variations are basically statistical fluctuations
randomly distributed. Thus the net effect of adding the MOS light
curves to the pn one results in an overall decrease of the level of
variability.
\begin{figure}
{\psfig{figure=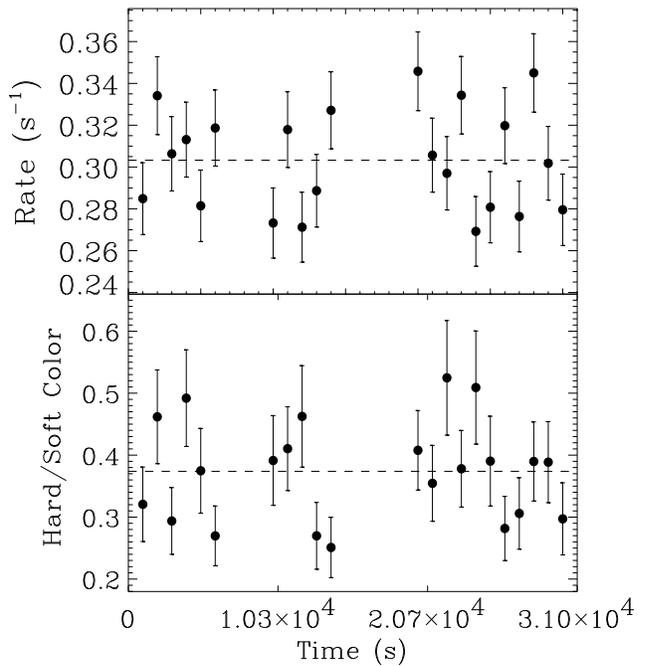,height=9cm,width=8.7cm,%
bbllx=37pt,bblly=7pt,bburx=485pt,bbury=460pt,angle=0,clip=}
\caption{EPIC pn light curves of the background-subtracted
count rate in the 0.3--10 keV band (top panel) and of the X-ray color 
2--10 keV/0.3--0.8 keV (bottom panel).
The extraction radius is 20\arcsec; time bins are 1000 s. The
dashed lines indicate the average values.}
\label{figure:lchr20}}
\end{figure}
In the following, therefore, we will use data only from the
pn camera (0.3--10 keV). 
To examine the influence of the extraction radius on the variability
we extracted pn light curves from circular regions of 
radii ranging between 10\arcsec\ and 30\arcsec. According to a $\chi^2$ test,
the flux variability is most pronounced when the extraction radius is 20\arcsec.
This choice represents a compromise between 
the necessity to increase the photon statistics and to minimize the contribution of the
extended emission, which is particularly important at soft energies (see 
Figure~\ref{figure:cha-ima}).
We then extracted energy-selected light curves in the
ultra-soft (0.3--0.8 keV), soft (0.8--2 keV), and hard (2--10 keV) energy bands. 
In Table~\ref{table:shortvar} we summarize the results of the short-term variability 
analysis obtained using time bins of 1000~s.

Figure~\ref{figure:lchr20} shows the background-subtracted light curve
from the EPIC pn in the total energy range 0.3--10 keV (top panel).  The
bottom panel shows a plot of the hardness ratio, defined as the
ratio of the 2--10 keV count rate to the 0.3--0.8 keV count rate, versus
time. Low-amplitude variations are present in both cases, confirming
our previous results (Paper I). Applying a $\chi^2$ test against the
hypothesis of constancy, we find that variability is highly
significant in both light curves: $\chi^2$ probabilities of 0.6\% and
3\% are found for the flux and hardness ratio, respectively.
\begin{figure}
{\psfig{figure=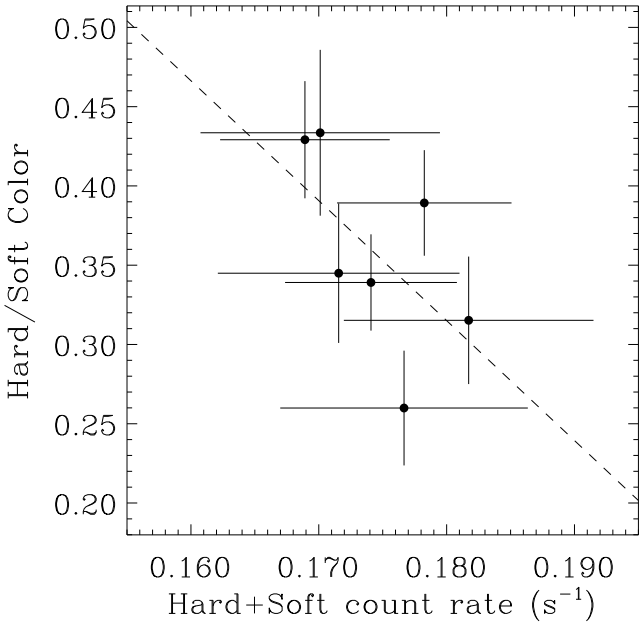,height=8.5cm,width=8.7cm,%
bbllx=37pt,bblly=65pt,bburx=415pt,bbury=430pt,angle=0,clip=}
\caption{EPIC pn X-ray color (2--10 keV/0.3--0.8 keV) versus the
count rate (2--10 keV + 0.3--0.8 keV); time bins are 4000 s. The
dashed line indicates the result of a linear least square fit,
showing a marginally significant ($P_c(r)=7.5\%$) anti-correlation.}
\label{figure:hrct}}
\end{figure}
The background count rate is as low as 5\% of the average source count rate
and, after removing two large flares, is consistent with the hypothesis 
of being constant, according to a $\chi^2$
test. To investigate the presence of possible systematic effects in 
the variability analysis, we extracted light curves from the only 
additional point-like source located on 
CCD4,
and from two of the brightest serendipitous sources located 
in the EPIC f.o.v. far from CCD edges
(RA=$12^h19^m50.1^s$, DEC=+$05^o$51\arcmin 04.75\arcsec, RA=$12^h19^m21.5^s$, 
DEC=+$05^o$42\arcmin 02.73\arcsec, and
RA=$12^h19^m04.7^s$, DEC=+$05^o$49\arcmin 04.8\arcsec, respectively).
All the sources are $\sim 3$ times
fainter than the central source, and
none of them is significantly variable according to $\chi^2$ testing.

The variability in different energy bands can be
characterized by means of the fractional variability parameter, $F_{\rm
var}$. The latter is a common measure of the intrinsic variability
amplitude relative to the mean count rate, corrected for the effect of
random errors, i.e.,
\begin{equation}
F_{\rm var}={(\sigma^2-\Delta^2)^{1/2}\over\langle r\rangle} ,
\end{equation}
where $\sigma^2$ is the variance, $\langle r\rangle$ the
unweighted mean count rate, and $\Delta^2$ the mean square value of
the uncertainties associated with each individual count rate. Using
different binning times ranging between 200 s and 1000 s, we calculated
$F_{\rm var}$ in the 0.3--0.8 keV band and in the $2-10$ keV band. For
any binning time, $F_{\rm var-ultrasoft}$ is larger than $F_{\rm
var-hard}$. For example, $F_{\rm var-ultrasoft} \sim
10\times10^{-2}$ for 200 s and $\sim 9\times10^{-2}$ for 1000 s, respectively, while
$F_{\rm var-hard} \sim5.5\times10^{-2}$ for 200 s and $\sim7\times10^{-2}$ for
1000~s bins.

Spectral variations are also present, as shown in Figure~\ref{figure:lchr20}. 
An interesting trend may be present when the hardness ratio is plotted
versus the total count rate. This is illustrated in
Figure~\ref{figure:hrct}, which shows the presence of a marginally
significant anti-correlation between the hardness ratios
and the total intensity. To quantify the degree of linear correlation
between $Hard/Soft$ and the mean count rate, we calculated the linear
correlation coefficient $r=-0.62$ and computed the chance probability
that a random sample of uncorrelated pairs of measurements would yield
a linear correlation coefficient equal or larger than $|r|$, finding
$P_c(r)=7.5\%$.

\section{Spectral Analysis}
\subsection{The X-ray continuum}
\begin{figure}[htb]
\psfig{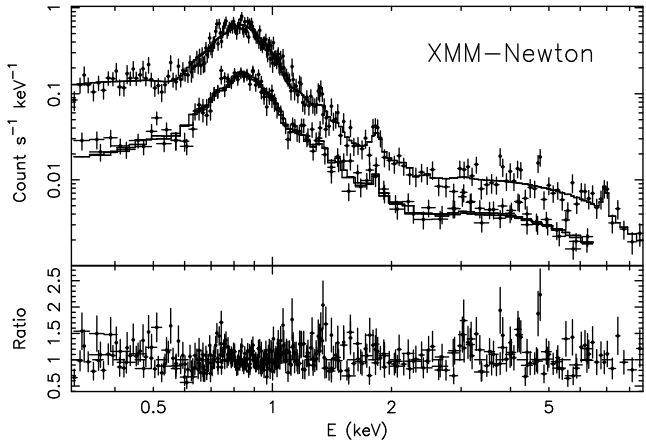}
\psfig{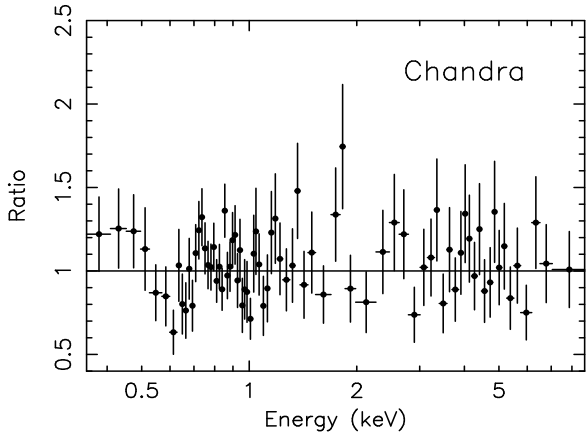}
\psfig{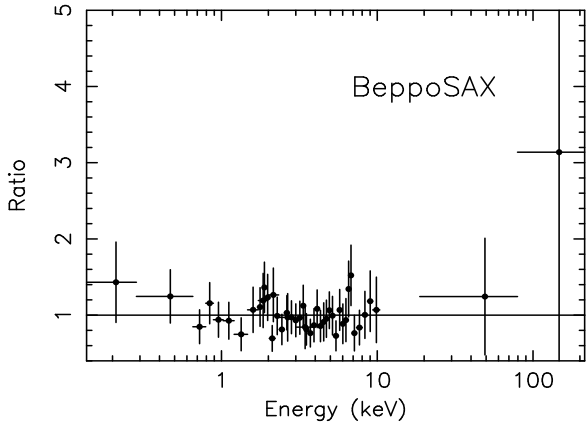}
\caption{{\it Top Panel:}
EPIC pn, MOS1, and MOS2 spectra of \object{NGC~4261} with the best-fit
model, consisting of a thermal component ({\tt apec} in {\tt XSPEC}),
plus a partially covered absorbed power law ({\tt zpcfabs powerlaw}),
plus a Gaussian line at $\sim$7 keV.
The same continuum model fits the \chandra\
and \sax\ data. {\it Middle Panel:} Residuals of the best-fit
continuum model to the ACIS data. {\it Bottom Panel:} Residuals of the
best-fit continuum model to the \sax\ LECS+MECS+PDS data. 
\label{figure:spectr}} 
\end{figure}

In Paper I, we presented an analysis of the EPIC pn spectrum from
0.3--10 keV. It was best-fit by a model consisting of a
thermal component related to the diffuse halo, dominating from
0.6--2 keV; a heavily absorbed power law above 2 keV, of nuclear
origin; and a second unabsorbed power law dominating the ultra-soft 
0.3--0.8 keV emission, with the photon index tied to that of the hard power
law.
The two power-law model mimics a partial-covering model. This
represents the physical situation of an absorber only partially
covering the nucleus, with the softer power law representing the
fraction of the flux which ``leaks through'' the patchy medium
(e.g., Holt et al. 1980). We therefore investigated the
combined spectra from the pn, MOS1, and MOS2 cameras (after checking the
consistency of each spectrum individually) using a thermal 
model plus a partially absorbed power law. 
Specifically, we fitted the \xmm\ EPIC data from 0.3--10 keV with the
following model: a thermal component, parameterized by \verb+apec+ in
\verb+XSPEC+, plus a partially absorbed power law, parameterized by
\verb+zpcfabs+\verb+(powerlaw)+, plus a uniform screen of absorbing gas with its
column density fixed to the Galactic value ($1.52\times10^{20}{~\rm cm^{-2}}$), acting
on all
components. The data with the best-fit model and the residuals are
shown in Figure~\ref{figure:spectr} (top panel), while the best-fit
parameters with their 90\% uncertainties 
are listed in Table 5. Line-like residuals are present at $\sim4.8$ keV;
however, adding a Gaussian line at that energy does not improve the fit
significantly. The 0.3--10 keV unabsorbed flux 
associated with the thermal component is $F_{\rm X,apec}=4.6\times10^{-13}
{~\rm erg~ cm^{-2}~ s^{-1}}$, whereas the flux associated with the power-law 
component is $F_{\rm X,pow}=11.2\times10^{-13}
{~\rm erg~ cm^{-2}~ s^{-1}}$.
\begin{table}
\caption{Spectral properties of NGC~4261 as seen by the \xmm\ EPIC cameras
and the \chandra\ ACIS-S3.}
\begin{center}
\begin{tabular}{lll}
\hline
\hline
\noalign{\smallskip}
Parameter & \xmm\ & \chandra\ \\
\noalign{\smallskip}
\hline
\noalign{\smallskip}
$kT$ (keV) & $0.65^{+0.01}_{-0.02}$ & $0.60^{+0.02}_{-0.02}$ \\ 
\noalign{\smallskip}
$Z~ (Z_\odot)$ & 1 ($>0.4$) & 1 \\ 
\noalign{\smallskip}
$N_{\rm H} ~(\rm 10^{22} cm^{-2})$  & $5.1^{+1.1}_{-1.2}$ & $6.8^{+2.5}_{-2.1}$ \\
\noalign{\smallskip}
CvrFract  & $0.81^{+0.06}_{-0.09}$ & $0.90^{+0.06}_{-0.07}$\\ 
\noalign{\smallskip}
$\Gamma$  & $1.46^{+0.17}_{-0.31}$ & $1.14^{+0.43}_{-0.38}$ \\ 
\noalign{\smallskip}
$E_{\rm line}$ (keV) & $6.99^{+0.08}_{-0.09}$ &  \\
\noalign{\smallskip}
$\sigma$ (keV) & $0.03^{+0.15}_{-0.03}$ &  \\
\noalign{\smallskip}
EW (eV) & $230^{+166}_{-134}$ &  \\
\noalign{\smallskip}
line flux ($10^{-6}{~\rm s^{-1}~cm^{-2}}$) & $1.7^{+1.3}_{-1.0}$ &  \\
\noalign{\smallskip}
$\chi^2/d.o.f.$& 400.91/359 & 76.15/71  \\ 
\noalign{\smallskip}
Counts & 11708 & 222  \\ 
\noalign{\smallskip}
$L_{\rm 0.3-2~keV}~(10^{40}{\rm erg~s^{-1}})$ &8.6 & 3.6 \\
\noalign{\smallskip}
$L_{\rm 2-10~keV}~(10^{40}{\rm erg~s^{-1}})$ &8.3 & 9.5 \\
\noalign{\smallskip}
 \hline
\end{tabular}
\end{center}
\end{table}
We find that the same continuum model fits the \chandra\ ACIS-S
data from 0.3--9 keV. The results of this spectral fit along with the intrinsic
luminosities are also listed
in Table 5, for direct comparison with \xmm, and the residuals are shown
in Figure~\ref{figure:spectr} (middle panel). 

The \xmm\ and \chandra\ best-fit parameters are consistent within the
90\% errors, with the exception of the temperature of the diffuse halo,
which is slightly
lower in the \chandra\ spectral fit. This is likely due to a shallow
temperature gradient in the circumnuclear region very close to the
central source. Indeed, dividing the inner 30\arcsec\ into six annular
regions and comparing the spatially-resolved \chandra\ spectra, we
find a shallow temperature gradient with $kT$ going from 0.6 to 0.8 keV
but with large uncertainties due to the limited statistics. The larger
$0.3-2$~keV luminosity measured by \xmm\ can be ascribed to the larger
extraction area (20\arcsec\ vs. 2\arcsec) used, meaning a stronger
contribution from the thermal component that actually peaks in the soft
X-ray range.

The \chandra\ data were previously analyzed by Chiaberge et
al. (2003). These authors found that a single absorbed power law
plus a thermal plasma best-fits the S3 data ($\chi^2/dof=90/69$). The
different spectral results can be partially ascribed to the different
processing of the data: 1) we use the latest released calibration files (CALDB
v2.18); 2) we apply {\tt ACISABS} to the ARF file to account for
the degradation of the QE due to molecular contamination; 3) we did
not convolve our spectral model with {\tt pileup} in {\tt XSPEC},
because, according to a test with PIMMS, the pile-up in
\object{NGC~4261} is less than 5\%. In any case, from the statistical
point of view our fit is
significantly better than that of Chiaberge et al. (2003).

We also tried to fit the data with more complex spectral models, e.g., 
{\tt pexrav} or the ionized-disk reflection model of Ross \& Fabian 
(1993). However, the fit did not improve.

The spectral analysis of the \sax\ data is inconclusive. 
The spectrum fitted by 
a thermal component ($kT=0.65\pm0.3$ keV) plus an absorbed power law
($N_{\rm H}\sim7\times 10^{21}{\rm~cm^{-2}}$, $\Gamma=1.3\pm0.3$) gives
a reduced $\chi^2$ of $\sim0.5$ for 49 d.o.f. Similar results are
obtained using more complex spectral models. The low value of $\chi^2_{\rm red}$
indicates that the fit
is dominated by the large statistical errors due to the poor photon statistics.
However, it is important to point out
that the PDS detected the source 
up to energies higher than 100 keV at more than a 3$\sigma$ confidence level.
The flux is $\sim 6\times10^{-12}{~\rm erg~cm^{-2}~s^{-1}}$ in the 10--100 keV range, 
implying an intrinsic luminosity of $\sim 6.5\times10^{41}{~\rm erg~s^{-1}}$.

\subsection{The Fe line}
\begin{figure}[h]
\psfig{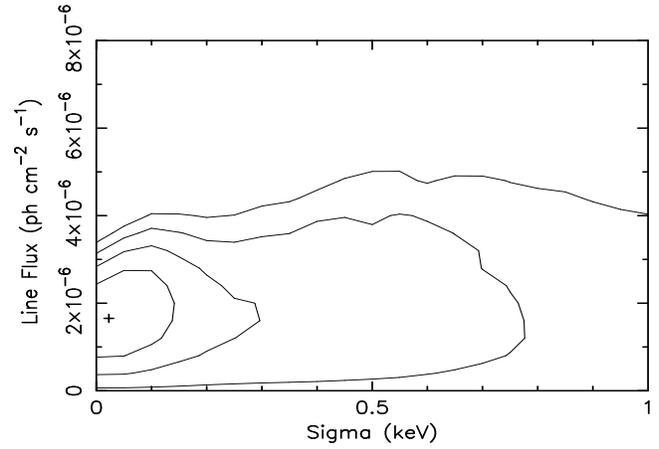}
\caption{Confidence contours (68\%, 90\%, 95\%, and 99\%) in the
$\sigma$ - line flux plane for the ionized iron line detected
with the \xmm\ EPIC pn camera. 
\label{figure:iron}} 
\end{figure}

Another important result of Paper I was the detection of an unresolved
Fe K line at $\sim$ 7 keV in the EPIC pn data.
Unfortunately, due to their limited sensitivity, the MOS data do not
provide any constraint on the Fe K line (Fig.~\ref{figure:spectr}). 
In Figure~\ref{figure:iron}, we
show the contours at 68\%, 90\%, 95\%, and 99\% confidence level
from the EPIC pn data for the line flux versus width. 
It is apparent that the line is detected at \gtsima 95\%
confidence and is not significantly resolved.
A broad line cannot be excluded. 
As mentioned in Paper I, no Fe line is detected in the \chandra\ ACIS
S3 data, with an EW upper limit of 320 eV. On the other hand, the \sax\ MECS data
show evidence for the Fe line, as indicated by the excess in the
residuals around 6--8 keV (see Fig.~\ref{figure:spectr}, bottom panel).
Adding a Gaussian model with the width fixed at the best-fit value from the EPIC pn data
to the best-fit MECS continuum gives a slight improvement in the fit
($\Delta\chi^2=2.4$ for two additional parameters). 
The line parameters are consistent with those from the EPIC pn, 
$E=6.7\pm0.7$ keV, $EW=304_{-304}^{+460}$ eV. 
Thus, we confirm the previous claim based on a 40 ks \asca\ exposure
that an Fe line is present in the spectrum of NGC~4261 (Terashima et
al. 2002; Sambruna et al. 1999). The line energy is consistent with
emission from Fe XXVI.
However, as discussed in Paper I, the origin of the line and the
nature of the reprocessor are still unclear.  A deeper exposure with
the EPIC pn is necessary to study the line profile in more detail. 
\section{Discussion}

\subsection{Summary of X-ray Results} 

We first summarize the main observational results that will be
relevant for the discussion of the nuclear properties of NGC~4261.
 
From the deprojected density profile (Eq. 2) and the spectral
analysis of the diffuse halo, 
one can place an upper limit on the accretion rate from the
interstellar medium onto the central black hole by applying the Bondi
theory of spherical accretion (Bondi 1952).  Specifically, the
Bondi accretion rate can be written (see, e.g., Di Matteo et al. 2003)
\begin{equation}
\dot M_{\rm Bondi}=4\pi R_{\rm
 A}^2 \rho_{\rm A} c_{\rm s},
\end{equation}
where $R_{\rm A}\simeq GM/c_{\rm s}^2$ is the accretion radius,
$c_{\rm s}\sim10^4 T^{1/2}{~\rm cm~s^{-1}}$ the sound speed, and
$\rho_{\rm A}$ the density 
at the accretion radius. According to the above
definition of $R_{\rm A}$ and using the best-fit gas
temperature, $kT\simeq 0.60$ keV (see $\S5$), the accretion radius of \object{NGC~4261}
is located at $\sim$0.4\arcsec$\simeq 55$ pc. Extrapolating the
density profile to this distance, we derive $\dot M_{\rm
Bondi}\simeq 0.04{~\rm M_\odot~yr^{-1}}$.  This value can be used to
calculate the expected accretion luminosity. Assuming a canonical
radiative efficiency of 10\%, we find $L_{\rm accr}\sim
2.5\times10^{44} {~\rm erg~s^{-1}}$.

From the best-fit spectral model, after correcting for the covering
fraction and the local absorption, the $0.3-10$ keV luminosity
associated with the nuclear non-thermal component is $L_{\rm
non-therm}\sim 1.2\times10^{41}{~\rm erg~s^{-1}}$. The estimate of the
X-ray luminosity can be further increased to $\sim
7\times10^{41}{~\rm erg~s^{-1}}$ by extending the spectrum up to 100 keV,
as observed with the PDS aboard \sax. It is worth noticing that the
non-thermal luminosity is $\sim350$ times smaller than the
accretion luminosity inferred assuming $\dot M_{\rm
accr}=\dot M_{\rm Bondi}$.

The main results of the timing analysis are 1) the presence of rapid
variability on time scales of a few ks; 2) the fact that the variability
is more pronounced in the ultra-soft 0.3--0.8 keV
energy band than in the hard 2--10 keV energy band; and 3) the possible
existence of an anti-correlation between the hardness ratio and the
total count rate.

\subsection{Properties of the VLBI Jet}

One of the goals of this paper is to determine the extent to which the jet
contributes to the nuclear X-ray emission from \object{NGC 4261}.
In this section, we investigate the properties of the
pc-scale jet focusing on its energetic requirements.

It is generally believed that the most important power content of extragalactic
jets is not in the form of internal random energy giving rise to observed
radiation, but in the form of kinetic power associated with particles and
magnetic fields. This requirement comes from the power budget of
extended radio structures, which need an average supply greater than
the luminosity generated by the jet at all scales (e.g., Rawlings \&
Saunders 1991; Ghisellini \& Celotti 2001). 

The total power transported by the flow is thus a fundamental property
for the discussion of jet energetics.  An estimate of the {\it
minimum} jet kinetic power can be obtained following the method of
Gliozzi, Bodo \& Ghisellini (1999).
This method assumes that the energy flux flowing
through the cross-section of the jet is carried by particles,
 $L_{\rm k,part}=\pi R^2 \Gamma^2\beta c n'(\langle \gamma
\rangle m_{\rm e} + m_+)c^2$, and
magnetic fields, $L_{\rm k,B}=\pi R^2 \Gamma^2\beta c U_{\rm B}$, 
where $R$ is the radius of the jet cross-section, $\Gamma$ the bulk
Lorentz factor, $\beta=v/c$, $n'$ the comoving particle density,
$\langle\gamma\rangle$ the mean Lorentz factor of the electrons, and
$U_B$ the magnetic energy density. The quantity $m_+$ is either the
proton mass $m_{\rm p}$ in the case of ``normal" plasma, or the
positron mass $\langle\gamma\rangle m_{\rm e^+}$ for $e^\pm$
pairs. A lower limit on the electron density $n'$ can be estimated
from the observed synchrotron emission $L_{\rm syn}$. Approximating
the VLBA jet as a cylinder of radius $R$ and length $h$, the number
density of leptons producing the observed radiation is
$n= {6L_{\rm syn}/(\langle\gamma^2\rangle  \delta^4 
\sigma_{\rm T}c B^2 h R^2}$),
where $\langle\gamma^2\rangle$ is averaged over the relativistic
electron distribution, $\delta=[\Gamma(1-\beta\cos\theta)]^{-1}$ is
the Doppler factor, and $\sigma_{\rm T}$ is the Thomson cross-section.
As the kinetic power associated with the particles has a dependence on
the magnetic field, $L_{\rm k,part}\propto B^{-2}$, the total kinetic
power can be minimized with respect to the magnetic field, $\partial
L_{\rm k,tot}/\partial B=0$. This yields a value of the magnetic field
$B_{\rm min}$ corresponding to the minimum power.

In blazars, where jets on pc-scales are unresolved due to the
small viewing angle, the jet parameters have to be inferred indirectly
by modelling the spectral energy distributions (SED) with specific
models (e.g., Tavecchio et al. 2000).
On the other hand, in
the case of \object{NGC~4261} the pc-scale jet parameters are obtained
directly from VLBA observations (Jones \& Wehrle 1997; Piner, Jones,
\& Wehrle 2001).  The jet spectral index is $\alpha=0.29\pm0.07$
($f\propto\nu^{-\alpha})$, and therefore the index of the electron
energy distribution is $p=2\alpha+1=1.58$. The jet velocity and
inclination to the line of sight are $\beta=0.46$ and $\theta=$ 63\deg,
respectively; as a consequence the bulk Lorentz factor is
$\Gamma=1.126 $ and the Doppler factor is $\delta=1.122$. The synchrotron
luminosity derived from radio data
is $L_{\rm syn}\simeq3\times10^{39} {~\rm erg~s^{-1}}$.
Note that a possible contribution from the UV jet is less than 10\% of
the radio luminosity (Chiaberge et al. 2003). A negligible contribution to the
synchrotron luminosity is expected from the X-rays on a twofold basis:
1) the steepness of the electron energy distribution, and 2) the fact
that the SED of \object{NGC~4261} has a secondary broad peak
located in the X-ray range (Lewis et al. 2003). From
the VLBA images the jet has transverse
and longitudinal dimensions of $\sim 2$ mas and $\sim 14$ mas, which
correspond to $R\simeq 8.8\times10^{17}$ cm and $h\simeq 6.9\times
10^{18}$ cm after deprojecting the longitudinal dimension.

The main uncertainties in deriving the jet kinetic luminosity are the
low- and high-energy cutoffs in the electron energy distribution,
$\gamma_{\rm min}$ and $\gamma_{\rm max}$, respectively.  An upper
limit $\gamma_{\rm max}\simlt 4\times10^5$ can be determined assuming that the
radio-to-UV emission is due to synchrotron emission from the same
electron population. This is a reasonable assumption for a mildly
relativistic jet, as inferred from the VLBA observations
of NGC~4261 (see Chiaberge et al. 2003 for a detailed discussion of
the UV jet). Given the steep electron
energy distribution, most of the energy is stored in low-energy
electrons, and therefore the low-energy cutoff plays a more important
role than $\gamma_{\rm max}$.  Since there is no direct and secure
way to set a lower limit on $\gamma_{\rm min}$, we carried out
the calculation of $L_{\rm kin}$ assuming $\gamma_{\rm min}$ ranging
between 1 and 30 (typical values inferred by blazar SED modelling;
e.g., Tavecchio et al. 2000), and we evaluated the influence of the
low-energy cutoff on $B_{\rm min}$ and $L_{\rm kin,min}$.

The magnetic field corresponding to the minimum power is $B_{\rm
min}=3.3$ mG, for a ``normal" plasma and $\gamma_{\rm min}=1$. A
somewhat lower value, $B_{\rm min}=2.4$ mG, is found assuming 
$\gamma_{\rm min}=30$. Similar values are found in the case of pair
plasma: $B_{\rm min}\simeq 2.5$ mG.
With a magnetic field strength equal to $B_{\rm min}$, the particle
kinetic power and the Poynting flux are nearly equal, and the total
power $L_{\rm k,tot}$ ranges between $3.7\times10^{40}{~\rm
erg~s^{-1}}$ and $1\times10^{40}{~\rm erg~s^{-1}}$ for normal and pair
plasmas, respectively. Slightly lower values are found assuming a low
energy cut-off of 30.  

The jet kinetic power must be multiplied by two to take into account
the contribution from the counter-jet.  It is worth noticing that
while the values of $B_{\rm min}$ are of the same order as the ones
derived for blazars,  
the kinetic luminosity is significantly lower.  This 
is due to the low value of $\Gamma$ inferred from the
VLBA observations; assuming $ \Gamma=10$, as in blazars, the kinetic
luminosity increases by several orders of magnitude ($L_{\rm
k,tot}\sim 2\times10^{44}{~\rm erg~s^{-1}}$) and becomes consistent 
with the values typically inferred for BL Lacs.

\subsection{Origin of the nuclear X-ray emission}
The strong correlation between radio and X-ray core fluxes found in
low-luminosity (Fabbiano et al. 1984; Canosa et
al. 1999) and high-luminosity (Worrall et al. 1994;
Hardcastle et al. 1998) radio galaxies has often been used to argue in favor of
a common origin from the unresolved base of the jet 
for the emission at the two wavelengths (e.g., Hardcastle
\& Worrall 1999 and references therein). More recently, a jet origin
for the X-rays was claimed for several low-power radio galaxies
observed with \chandra\ (Pellegrini et al. 2003; Fabbiano et al. 2003;
Chiaberge et al. 2003).

We note the following:\\ 1) While the strong correlation between radio 
and X-ray fluxes does suggest a physical connection, it does
not necessarily imply a common jet origin of the two fluxes.  Indeed,
accretion onto compact objects and relativistic jets seem to be
correlated phenomena (e.g. Begelman, Blandford \& Rees 1984). 
 As a consequence,
at some level, a correlation between the jet and the  
disk-corona\footnote[2]{For simplicity, we
use the term ``disk corona" to indicate any type of accretion flow at work
around the black hole.} flux is expected.

2) The claim of a jet
origin for the X-rays, in addition to the radio-X-ray correlation, is
based on spectral results. However, a common problem
affecting the study of AGN is spectral degeneracy: the same
spectrum can be equally well described by quite different spectral
models. This is exacerbated in the case of low-power radio
galaxies, due to their low signal-to-noise ratio X-ray spectra.  For
example, in the case of \object{NGC~4261}, the photon index
$\Gamma\sim 1.5\pm0.3$ is consistent with inverse Compton emission from a jet
but also with an advection dominated accretion flow 
(ADAF) model. It is worth noticing that Fabbiano et al. (2003)
fit satisfactorily the radio-to-X-ray SED of the low-luminosity radio galaxy
\object{IC~1459} with a jet-dominated model. 
However, as already highlighted by Di Matteo et al.
(2001a) for several nearby galaxies, this simply indicates that analytical 
ADAF models are unable to account for the high radio flux and that an additional
contribution, possibly from a small-scale radio jet, is required. It does not
necessarily imply that a pure jet model is the solution. In fact, this latter model,
tentatively applied to higher quality and much broader bandpass energy spectra,
is far from being widely accepted yet (e.g., Zdziarski et al. 2003).
 
In order to break the spectral degeneracy, one needs to exploit
additional information coming, for example, from temporal
analysis. Relying upon timing and spectral variability results
from long monitoring campaigns of two powerful broad-line radio
galaxies (BLRG; \object{3C~390.3} and \object{3C~120}), Gliozzi, Sambruna \&
Eracleous (2003) concluded that the X-ray emission 
is not dominated by jets. Both BLRGs display
typical X-ray behavior observed in Seyfert galaxies, namely a softer
X-ray spectrum with increasing flux (e.g.,
Petrucci et al. 2000; Vaughan \& Edelson 2001; Papadakis
et al. 2002) and a larger variability amplitude in soft X-rays (Nandra
et al. 1997; Markowitz \& Edelson 2001).
Moreover, jet-dominated sources, i.e., blazars, typically show the opposite
temporal and spectral behavior: the X-ray spectrum hardens when the
flux increases and the variability amplitude increases toward higher energies
(see, e.g., Zhang et al. 1999; Fossati et al. 2000).
In the case of \object{NGC~4261}, the
temporal and spectral variability behavior similar to that shown by \object{3C~390.3} and
\object{3C~120} (although with a lower statistical significance) is suggestive evidence 
that most of the X-rays originate from the disk-corona system.
This favors the conclusion that the jet is not the dominant
mechanism producing the X-rays in the nucleus of NGC~4261.

This conclusion is independently confirmed if we compare the
jet kinetic power with the X-ray luminosity. Let us assume that most of the
X-rays originate from the jet; this would lead to the unphysical
conclusion that the radiative power of the jet is more important than
its kinetic power. This would be at odds with the energetic
requirements of the lobes (see \S6.2). One can object that
the estimated $L_{\rm kin}$ represents only the minimum value for the 
kinetic power of the jet, and that, if the jet is not in equipartition,
this value can significantly increase. However, one must keep in mind that
1) moderate departures from equipartition
do not increase $L_{\rm kin}$ by orders of magnitude (Ghisellini \& Celotti 2001);
2) typical values for the
kinetic-to-radiative luminosity ratio in radio-loud AGNs are estimated to be 
$\sim 10^2-10^4$ (see, e.g., Celotti \& Fabian 1993).

\subsection{Nature of the accretion flow}
Before starting the discussion on the nature of the accretion flow,
it is necessary to emphasize the distinction between radiatively
inefficient accretion flow (RIAF) and ADAF
models. The former class describes the properties of rotating accretion
flows where very little of the accretion energy is radiated away, whereas 
ADAF models are simple analytical models for the dynamics of RIAF models.
The most relevant difference is that ADAF models predict that the rate at which
gas accretes onto the black hole is comparable to the Bondi accretion rate; therefore,
the low luminosity in ADAF models is just due to very low radiative efficiency.
On the other hand, time-dependent numerical simulations of RIAF models indicate
that  $\dot M_{\rm accr}\ll \dot M_{\rm Bondi}$, implying that the observed low luminosity
is also due to a low accretion rate, rather than just a low efficiency.
For a more detailed discussion of RIAF models see, e.g., Quataert (2003).

Recently, \chandra\ studies of the nature of the accretion flows
in several giant elliptical galaxies have returned some controversial results 
on the relation between $\dot M_{\rm Bondi}$ and $\dot M_{\rm accr}$.
For example, Di Matteo et al. (2001b, 2003) question
whether $\dot M_{\rm Bondi}$ is a reliable estimate of $\dot M_{\rm accr}$
and whether the ADAF model is a viable solution for
\object{NGC~6166} and \object{M87}. In this latter case,
if feedback effects
from the base of the jet on the accretion flow are taken into account,
$\dot M_{\rm accr}$ can be as low as $10^{-6} \dot M_{\rm Bondi}$.
Also Loewenstein et al. (2001) conclude that
$\dot M_{\rm accr}$ must be significantly smaller than $\dot M_{\rm
Bondi}$ to find X-ray fluxes in agreement with the values observed by
\chandra\ for \object{NGC~1399}, \object{NGC~4472}, and
\object{NGC~4636}. On the other hand, Pellegrini et al. (2003) and Fabbiano et
al. (2003), investigating respectively \object{IC~4296} and
\object{IC~1459}, conclude that $\dot M_{\rm Bondi}$ can be a
reliable estimate of $\dot M_{\rm accr}$. However,
the authors exclude the ADAF scenario on the basis of SED
considerations, in particular the high radio-to-X-ray flux ratio,
favoring instead a jet-dominated model.

In the case of \object{NGC~4261}, on the basis of the \chandra\ data alone
and according to the Bondi theory, the first conclusion is that the black
hole is not fuel-starved ($\dot M_{\rm Bondi}\sim0.04{~\rm
M_\odot~yr^{-1}}$).  Given that the expected accretion luminosity
$L_{\rm accr}\sim 2.5\times 10^{44}{~\rm erg~s^{-1}}$ (assuming a
canonical efficiency of 10\%) is much higher than the disk-corona luminosity,
we would be tempted to conclude that accretion is radiatively
inefficient. Indeed, using the results derived by Loewenstein et
al. (2001) for \object{NGC~1399} (which has $M_{\rm BH}$ and $M_{\rm Bondi}$  
similar to those of \object{NGC~4261}), by applying the basic ADAF model, we can conclude
that the 2--10~keV luminosity of \object{NGC~4261} is roughly
consistent with $L_{\rm ADAF}$.  Therefore, disregarding the problem
of the high radio-to-X-ray flux ratio, possibly solved by ascribing to the base 
of the jet most of the core radio emission (see e.g., Di Matteo
et al. 2001a), at the zeroth order an ADAF (+ radio jet) seems to be
an acceptable solution.

An independent way to test whether $\dot M_{\rm Bondi}$ is a reliable
estimate of $\dot M_{\rm accr}$ is based on a comparison between the
accretion rate and outflow rate, $\dot M_{\rm out}$, which can be
derived from the kinetic energy of the jet: $L_{\rm k}=(\Gamma
-1)\dot M_{\rm out}c^2$. Assuming $\dot M_{\rm Bondi}$=$\dot M_{\rm accr}$
and the value of the kinetic power derived in $\S 6.2$, the conclusion
would be that $\dot M_{\rm out}\sim 10^{-4}\dot M_{\rm accr}$. This value 
seems low considering that studies of the
jet-disk luminosities of the BL Lac class (the parent population of
FR~I according to the unification models; e.g., Urry \& Padovani 1997) 
indicate that the channelling of
matter from the accretion flow into the jet outflow seems to be
particularly efficient. 
For example, Maraschi \& Tavecchio (2003) find 
that for BL Lacs $L_{\rm disk}/L_{\rm k}\sim 10^{-3}-10^{-4}$.
Applying this result
to \object{NGC~4261} with $L_{\rm disk}=\eta \dot M_{\rm accr} c^2$,
$M_{\rm accr}=M_{\rm Bondi}$, and 
$L_{\rm k}$ given by the above formula, we obtain an unphysically low radiative 
efficiency, $\eta\sim10^{-8}-10^{-9}$.

Further independent evidence that 
the $\dot M_{\rm out}/\dot M_{\rm accr}$
derived for \object{NGC~4261} is unreasonably small comes from MHD simulations
that show that most of the inflowing flow is lost in a magnetically driven wind
(e.g., Stone \& Pringle 2001; Hawley \& Balbus 2002).
As a consequence, energetic considerations favor a RIAF model, where 
the intrinsically low
luminosity of \object{NGC~4261} is explained in terms of low accretion rate
coupled with a moderately low radiative efficiency. 

Additional important pieces of
information can come from the \xmm\ observations, in particular
from the temporal analysis.
Even though, historically,
rapid X-ray variability in AGN has been associated with standard
accretion disks, ADAF models can also account for such variability
if the X-rays are produced by Compton process in the inner region
of the accretion flow.
The rapid flux variations detected from \object{NGC~4261}
indicate that such
variability can actually occur also in low-power AGNs and that RIAF models should
account for this temporal behavior consistently. Indeed, the rapid X-ray flares 
observed from Sgr~${\rm A^\ast}$ have been recently interpreted by Yuan et al. (2003)
in the framework of radiatively inefficient accretion flow models. 

\section{Summary  and Conclusions}

We have presented \xmm, \chandra, and \sax\ observations of the
low-power FR~I radio galaxy \object{NGC~4261}, host of a supermassive
black hole and an optical LINER. The main findings can be
summarized as follows:\\
$\bullet$ Several spatial components are present in the X-ray image: an
unresolved nuclear region, present in all energy bands; a diffuse hot
halo, dominating the 0.6--2~keV emission; a double-sided kpc-scale jet;
and several point-like sources. The jet and the point sources give a
negligible contribution to the X-ray flux in the \xmm\ aperture.\\
$\bullet$ The surface brightness of the extended circumnuclear emission is well fitted
with a $\beta$-model ($\beta\simeq0.53$, $r_c\simeq1.40$\arcsec). Combining these
results with the spectral analysis ($kT\sim 0.6-0.65$ keV) and using the Bondi
theory, an accretion rate of $\dot M\simeq 0.04 {~\rm M_\odot~yr^{-1}}$
is found, corresponding to an accretion luminosity of $L_{\rm accr}\sim 2.5\times10^{44}
{~\rm erg~s^{-1}}$ for a radiative efficiency of 10\%.\\
$\bullet$ The spectrum of the nucleus is well described from 0.3--10 keV by a
power law with $\Gamma\simeq1.5$, partially covered by cold gas
(CvrFract$\simeq 0.8$, $N_{\rm H}\simeq 5\times 10^{22}{~\rm
cm^{-2}}$), plus a highly ionized unresolved iron line at $\sim 7$
keV. The presence of a line is confirmed by \sax, which also detects
at the 3$\sigma$ level high-energy emission up to 100--150 keV.\\ 
$\bullet$ The total non-thermal luminosity in the 0.3--10 keV band is
1.2$\times 10^{41}{~\rm erg~s^{-1}}$ ($\sim 7\times 10^{41}{~\rm erg~s^{-1}}$
between 0.3 and 100 keV), which is more than 2 orders of magnitude lower than 
the expected accretion luminosity.\\ 
$\bullet$ Significant flux and spectral variations of the nuclear X-ray emission
were detected with EPIC.  The variability is more pronounced
in the soft band (0.3--0.8 keV), and the hardness ratio appears to be
anti-correlated with the count rate. We conclude that most of the nuclear X-ray 
emission is related to the accretion flow and that the base of the jet is {\it
not} the dominant source of X-rays from the nucleus of NGC~4261.\\
$\bullet$ Energetic arguments based on the jet kinetic power support this conclusion
and suggest that the Bondi accretion rate overestimates (presumably by orders 
of magnitude) the actual accretion rate onto the black hole. Therefore, the low
X-ray luminosity of \object{NGC~4261} can be explained in terms of low accretion rate
coupled with a moderately low radiative efficiency, not just a very low efficiency
as hypothesized by ADAF models.
These results suggest some general conclusions. Although great
progress in the study of accretion in AGN is made by comparing spectral models
with the data, spectral results alone cannot unequivocally constrain
the origin of the X-rays and the nature of the accretion onto the
central black hole. This is especially true for low-power radio galaxies,
whose low signal-to-noise spectra can be acceptably fitted with
completely different physical models. Spectral analysis is further
complicated by the presence of several distinct components
contributing to the total X-ray emission.  
Major progress in the understanding of accretion phenomena can only be
reached by combining spectral results with independent information 
from spatial and timing analyses.  To this aim, constraints from
{\it both} \chandra\ and \xmm\ are needed.
\vskip 0.4 cm
\noindent{\centerline{\bf APPENDIX}}
\noindent{\centerline{\bf  Serendipitous sources in the Chandra field}}
\vskip 0.2 cm
\noindent The \chandra\ ACIS-S image of \object{NGC~4261} in the 0.3--10 keV energy range 
(Figure~\ref{figure:cha-ima}, bottom right panel)
reveals the presence of several off-nuclear point sources. 
We used the \verb+CIAO+ tool
{\tt wavdetect} to search for serendipitous X-ray sources in the
f.o.v
\begin{table}[hb]
\caption{Off-nuclear X-ray point-like sources}
\begin{center}
\begin{tabular}{lllll}
\hline
\hline
\noalign{\smallskip}
Source & RA       & DEC & Counts ${\rm ^a}$ & D ${\rm ^b}$  \\
      & (J2000)   & (J2000)   &  [0.35--9 keV]      & [\arcsec] \\
\noalign{\smallskip}       
\hline
\noalign{\smallskip}
\noalign{\smallskip}
A & $12^h19^m24.9^s$ & +$05^o$49\arcmin32\arcsec & $15\pm4$ & 26  \\
\noalign{\smallskip}
\hline
\noalign{\smallskip}
B & $12^h19^m24.0^s$ & +$05^o$49\arcmin27\arcsec & $33\pm7$ & 13\\
\noalign{\smallskip}
\hline
\noalign{\smallskip}
C & $12^h19^m23.5^s$ & +$05^o$49\arcmin36\arcsec & $81\pm10$ & 7.5\\
\noalign{\smallskip}
\hline
\noalign{\smallskip}
D & $12^h19^m22.7^s$ & +$05^o$49\arcmin43\arcsec & $22\pm5$ & 15.5\\
\noalign{\smallskip}
\hline
\noalign{\smallskip}
E & $12^h19^m20.3^s$ & +$05^o$49\arcmin27\arcsec & $15\pm4$ & 34 \\
\noalign{\smallskip}
\hline
\noalign{\smallskip}
\end{tabular}
\end{center}
${\rm ^a}$ For comparison, the number of counts found in the
central 2\arcsec\ source region is $2254\pm 47$.\\ 
${\rm ^b}$ Distance in arcsecond from \object{NGC~4261}.
\end{table}  
The algorithm returns a list of elliptical regions that
define the positions and the shapes of the detected sources (see Table 4).
We next used the coordinates from {\tt wavdetect} and its associated error
regions to search for optical counterparts on ESO archival plates and \hst\ images.
However, none of the five serendipitous sources has an apparent optical counterpart.
Source $C$ has
enough counts for temporal and spectral analyses. The spectrum, which
extends only in the 0.5--1.5 keV range, is
well fitted by a simple power law ($\Gamma=2.3\pm0.6$) absorbed
by Galactic $N_{\rm H}=1.52\times 10^{20}{~\rm cm^{-2}}$, with an
absorbed flux of $f_{0.5-1.5 \rm keV}=8\times10^{-15}{~\rm
erg~cm^{-2}~s^{-1}}$. The light curve is consistent with the source flux
being constant ($P_{\chi^2}> 99.9\%$). 
\vskip 0.5 cm

\begin{acknowledgements} 
We are grateful to  T. Cheung, M. Chiaberge, T. Di Matteo, D. Donato, M. Eracleous, J. Krolik, 
Alex Rinn, and F. Tavecchio for helpful discussions.
Financial support from NASA LTSA grants NAG5-10708 (MG, RMS)
and NAG5-13035 (WNB) is gratefully acknowledged. Funds
were also provided by NASA grant NAG5-10243 (MG, RMS) and by the Clare
Boothe Luce Program of the Henry Luce Foundation (RMS).
\end{acknowledgements}

\end{document}